\documentclass[11pt,a4paper]{article}
\usepackage{jheppub_kim}

\usepackage{pdflscape}
\usepackage{amsmath}
\usepackage{amssymb}
\usepackage{dcolumn}
\usepackage{bm}
\usepackage{color}
\usepackage{epsfig}
\usepackage{amsfonts}
\usepackage{graphicx}
\usepackage{subfigure}
\usepackage{dcolumn}

\newcommand{\be}{\begin{equation}}
\newcommand{\ee}{\end{equation}}
\newcommand{\bea}{\begin{eqnarray}}
\newcommand{\eea}{\end{eqnarray}}

\def\be{\begin{equation}}
\def\ee{\end{equation}}
\def\bea{\begin{eqnarray}}
\def\eea{\end{eqnarray}}

\begin{document}

\title{Logarithmic corrected Polynomial $f(R)$ inflation mimicking a cosmological constant}

\author[a]{J. Sadeghi}
\author[b]{B. Pourhassan}
\author[c]{A.S. Kubeka}
\author[d]{M. Rostami}

\affiliation[a]{Department of Physics, University of Mazandaran, P.O.Box 47416-95447, Babolsar, Iran}
\affiliation[b]{School of Physics, Damghan University, Damghan, Iran}
\affiliation[c]{Department of Mathematical Sciences, University of South Africa, Science Campus, P.O.Box 392, Florida, South Africa}
\affiliation[d]{Department of physics, Sari Branch, Islamic Azad University, Sari, Iran}

\emailAdd{pouriya@ipm.ir}
\emailAdd{b.pourhassan@du.ac.ir}
\emailAdd{kubekas@unisa.ac.za}
\emailAdd{M.Rostami@iauamol.ac.ir}

\abstract{In this paper, we consider an inflationary model of $f(R)$ gravity with polynomial form plus logarithmic term. We calculate some cosmological parameters and compare our results with the Plank 2015 data. We find that presence of both logarithmic and polynomial corrections are necessary to yield slow-roll condition. Also, we study critical points and stability of the model to find that it is a viable model.}

\keywords{Modified Gravity; Inflation; Plank Data.}

\maketitle

\section{Introduction}
Dark energy may be origin of the late-time cosmic accelerated expansion which observed by type Ia supernovae (SNIa) \cite{Riess:1998cb,Perlmutter:1998np}. Dark energy is a mystery thing with negative pressure \cite{Huterer}. Observational data of SNIa \cite{Riess:2004a}, CMB \cite{CMB} and BAO \cite{BAO} confirmed that most of universe filled with the dark energy. Proposing a successful model to describe dark energy is one of the important topics in theoretical physics and cosmology.\\
One of the first models is the cosmological constant which is non-dynamical model. So, some people proposed dynamical dark energy models such as scalar field models \cite{Liddle:1998xm,Guo:2006ab,Khurshudyan:2014a,Dutta:2009yb,Chiba,Armendariz,Onemli:2004mb,
Saridakis:2008fy,Zhao:2006mp,Cai:2009zp}.\\
There are also another class of modified matter models based on exotic fluid so called Chaplygin gas and its extensions \cite{P89,P92,P93,P96,P97,P100-1,P100-2,P102,P103,P1002,P1003,PLB636(2006)86,0812.0618,P26 1103.4842,IJTP52(2013)4583,PLB646(2007)215,P59 1012.5532,P60 1102.1632,ASS341(2012)689}, which is indeed a way to unification of dark matter and dark energy.\\
The other class of dynamical dark energy models are based on modification of the general relativity, which is called $f(R)$ or $F(R)$ gravity, where $F$ is a function of the Ricci scalar and $f=R+F$ \cite{Capozziello1,Capozziello2,Capozziello3,Carroll,Nojiri}. This is indeed an alternative way for introducing dark energy to have cosmic acceleration. Now, an important point is constructing a model in agreement with observational data. An easy way to do that  is to find $f(R)$ or $F(R)$. Polynomial $f(R)$ model is one of the possible way \cite{Huang}, and is a general model with $R^{2}$ correction. The higher derivative corrections of the ordinary model inspired by string theory is quite generic. However, as pointed out by the Ref. \cite{Huang}, other correction terms expected by a quantum gravity theory like string theory. For example, it is also possible to consider logarithmic corrected term. The ordinary function without polynomial and logarithmic terms ($f(R) = R+\alpha R^{2}$) \cite{42, 43} may be useful to study neutron stars with a strong magnetic field \cite{44}. Logarithmic corrections may be add to consider the effect of gluons in non-flat space-time \cite{45}. Recently, a cosmological model based on the logarithmic corrected $F(R)$ constructed \cite{46}.\\
In this paper, we would like to consider both polynomial form of $f(R)$ and a logarithmic term, to investigate the spectral index and tensor-to-scalar ratio comparing with recent plank data \cite{2015}. According to the recent Planck data the constraint on spectral index is given by $n_{s}=0.968\pm0.006$ consistent with the 2013 Planck data. Moreover, Planck 2015 plus BKP B-mode data give a constraint on the tensor-to-scalar ratio as $r<0.09$ \cite{2015}. Finally the number of e-foldings between the end of inflation
and our present day is $50\leq N\leq60$. Indeed, present paper in extension of \cite{Huang} and \cite{46} using more recent observational data. We will show that, presence of logarithmic plus polynomial corrections help to obtain results in more agreement with observational data.

\section{The polynomial plus logarithmic inflation model}
The inflationary expansion of the universe may described by the following action \cite{Huang},
\begin{equation}\label{M1}
S = \int d^4 x \sqrt{-g}
\left[\frac{1}{2\kappa^2} f(R)+ \mathcal{L}_{m} \right ],
\end{equation}
where $\kappa^{2}= 8\pi G = M^{-2}_{pl}$, and $M_{pl}$ is the reduced Planck mass, and $\mathcal{L}_{m}$ is the matter Lagrangian density. The main purpose of this paper is to study the cosmological parameters
describing inflation and the evolution of the universe. We consider $f(r)$ as follow,
\begin{equation}\label{f1}
f(R) = R+\alpha R^{2}+\beta R^{n}+\gamma R^{2}\ln{\gamma R},
\end{equation}
where $\alpha$, $\beta$ and $\gamma$ are arbitrary constant, and $n>2$. The case of $\alpha\neq\beta\neq0$ and $\gamma=0$ yield to the model considered by the Ref. \cite{Huang}. When $\beta=0$ the model reduced to the Starobinsky model \cite{Starobinsky}. The special case of $\gamma=0$ and $n=4$ has been studied by the Ref. \cite{Saidov}. The function $f(R)$ satisfies the conditions $f(0)= 0$, corresponding to the flat space-time without cosmological constant. We will compare our results with recent observation data to show that two last terms of (\ref{f1}) are necessary.\\
From the equation (\ref{f1}) we obtain,
\begin{eqnarray}\label{M2}
f^{\prime}(R) &=& 1 + (2\alpha+\gamma) R + 2 \gamma R \ln \beta R+n\beta R^{n-1}, \nonumber\\
f^{\prime \prime}(R) &=& 2\alpha+\lambda + 2 \gamma \ln \beta R+n(n-1)\beta R^{n-2},
\end{eqnarray}
The function $f(R)$ obeys the quantum stability condition $f^{\prime \prime}(R) > 0$ for $ \alpha > 0$, $ \beta >
0$, and $\gamma>0$. This ensures the stability of the solution at high curvature.
It follows from the equation (\ref{M2}) that the condition of classical stability
$f^{\prime}(R) > 0$ leads to,
\begin{equation}\label{M3}
1 + \left(2\alpha+\gamma + 2\gamma \ln \gamma R \right) R+n\beta R^{n-1} > 0.
\end{equation}

First of all we consider constant curvature solutions of the equations obtained from the action in the equation (\ref{M1}) without matter ($\mathcal{L}_{m}=0$). The governing equation is
given by \cite{B},
\begin{equation}\label{CCC1}
2 f(R) - R f^{\prime}(R) = 0,
\end{equation}
hence,
\begin{equation}\label{CCC2}
(2-n)\beta R^{n-1}-\gamma R+1=0.
\end{equation}
For the case of $\beta=0$ we have $R=\frac{1}{\gamma}$. In the case of $\beta\neq0$ we have an equation of order $n-1$. For instance, the case of $n=3$ yields to,
\begin{equation}\label{n=3}
R=-\frac{\gamma\pm\sqrt{\gamma^{2}+4\beta}}{2\beta}.
\end{equation}
There is also another condition,
\begin{equation}\label{CCC3}
\frac{f^{\prime}(R)}{f^{\prime \prime}(R)}> R,
\end{equation}
which simplifies to,
\begin{equation}\label{CCC4}
1-2\gamma R-n(n-2)\beta R^{n-1}> 0,
\end{equation}
which implies flat space-time ($R=0$) is stable. However, both conditions (\ref{CCC1}) and (\ref{CCC3}) satisfied for $n\geq3$ and $R>0$.

\section{The scalar field and potential}
\begin{figure}[h!]
 \begin{center}$
 \begin{array}{cccc}
\includegraphics[width=50 mm]{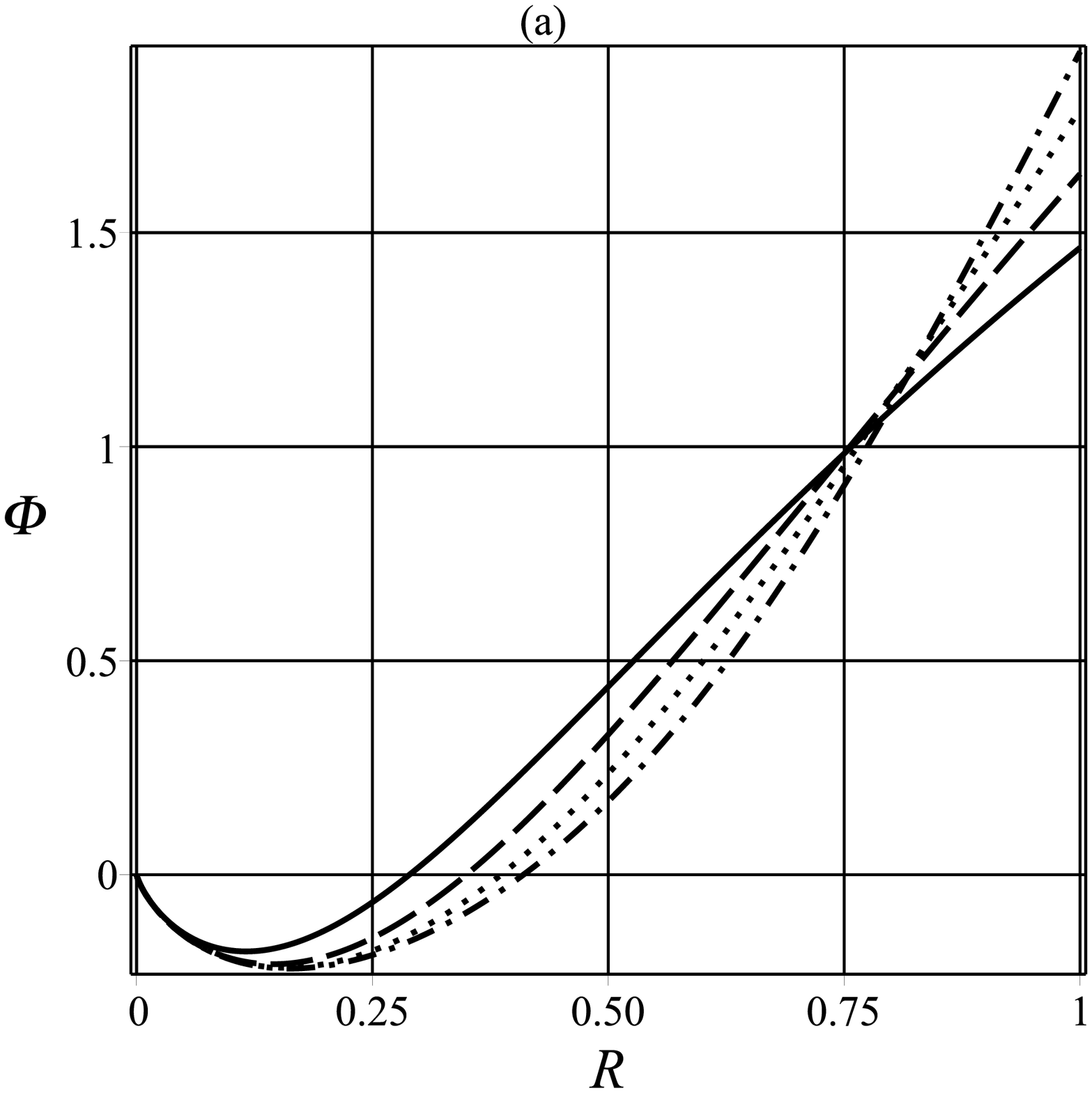}\includegraphics[width=50 mm]{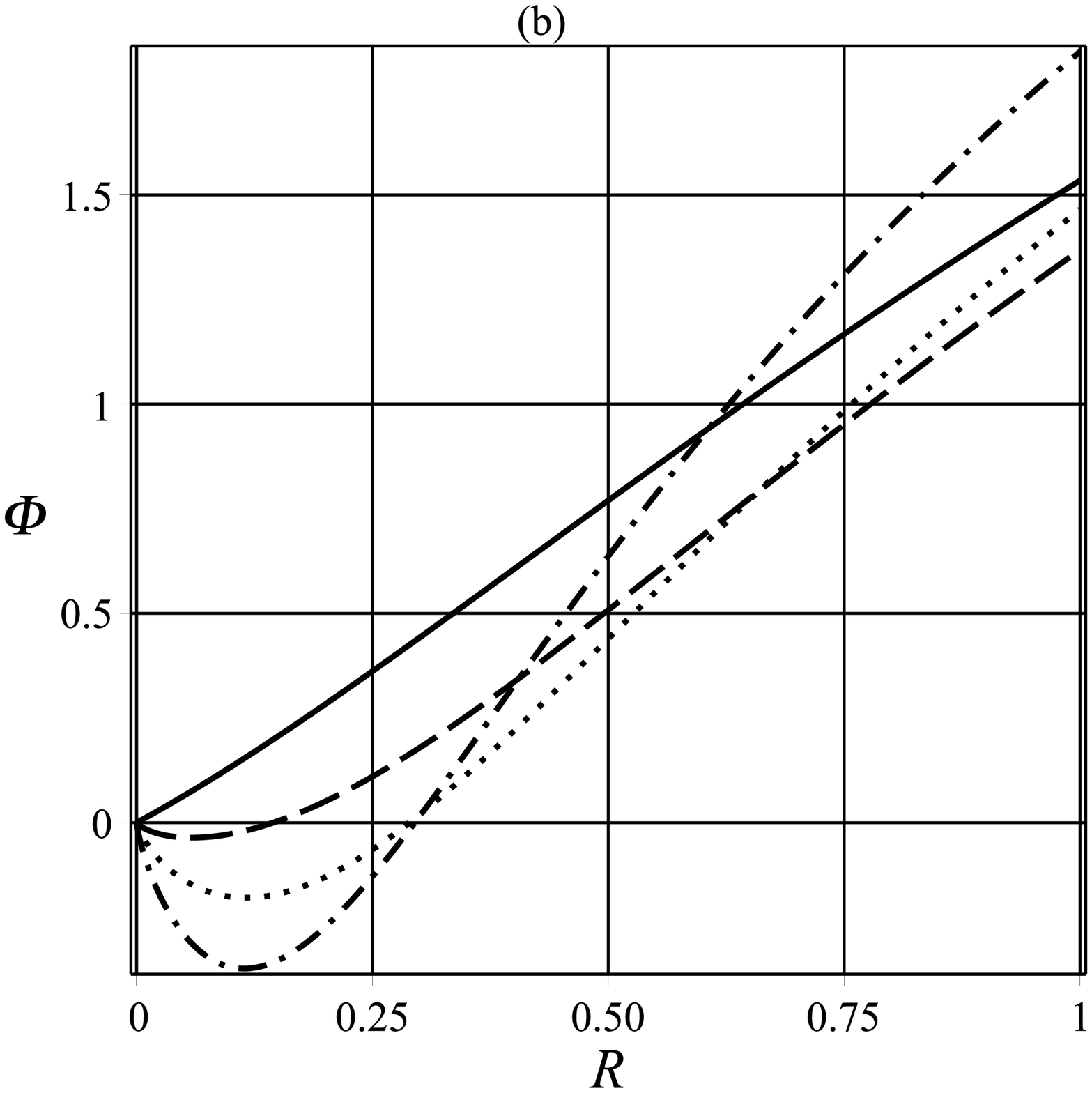}\includegraphics[width=50 mm]{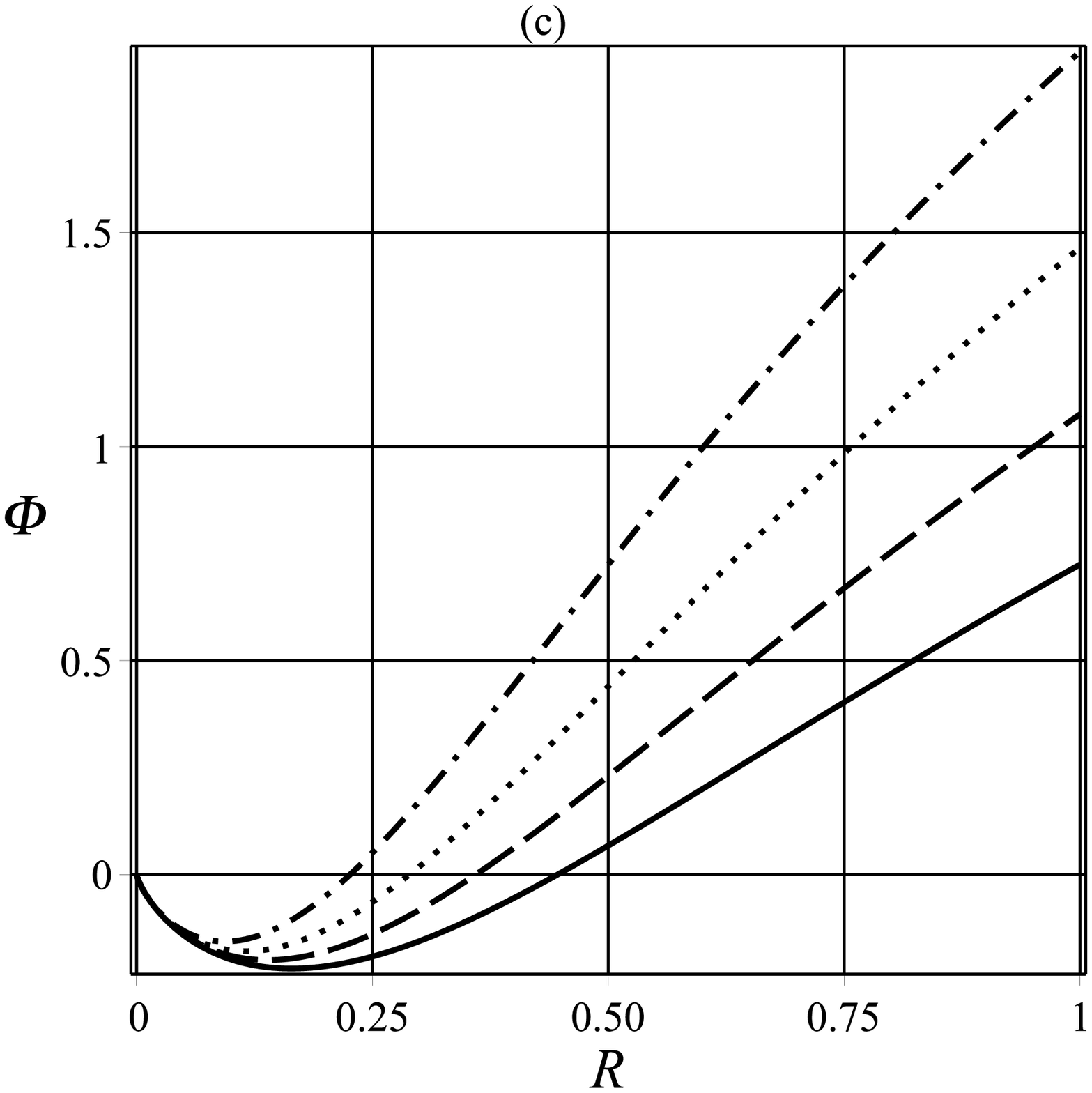}\\
\includegraphics[width=50 mm]{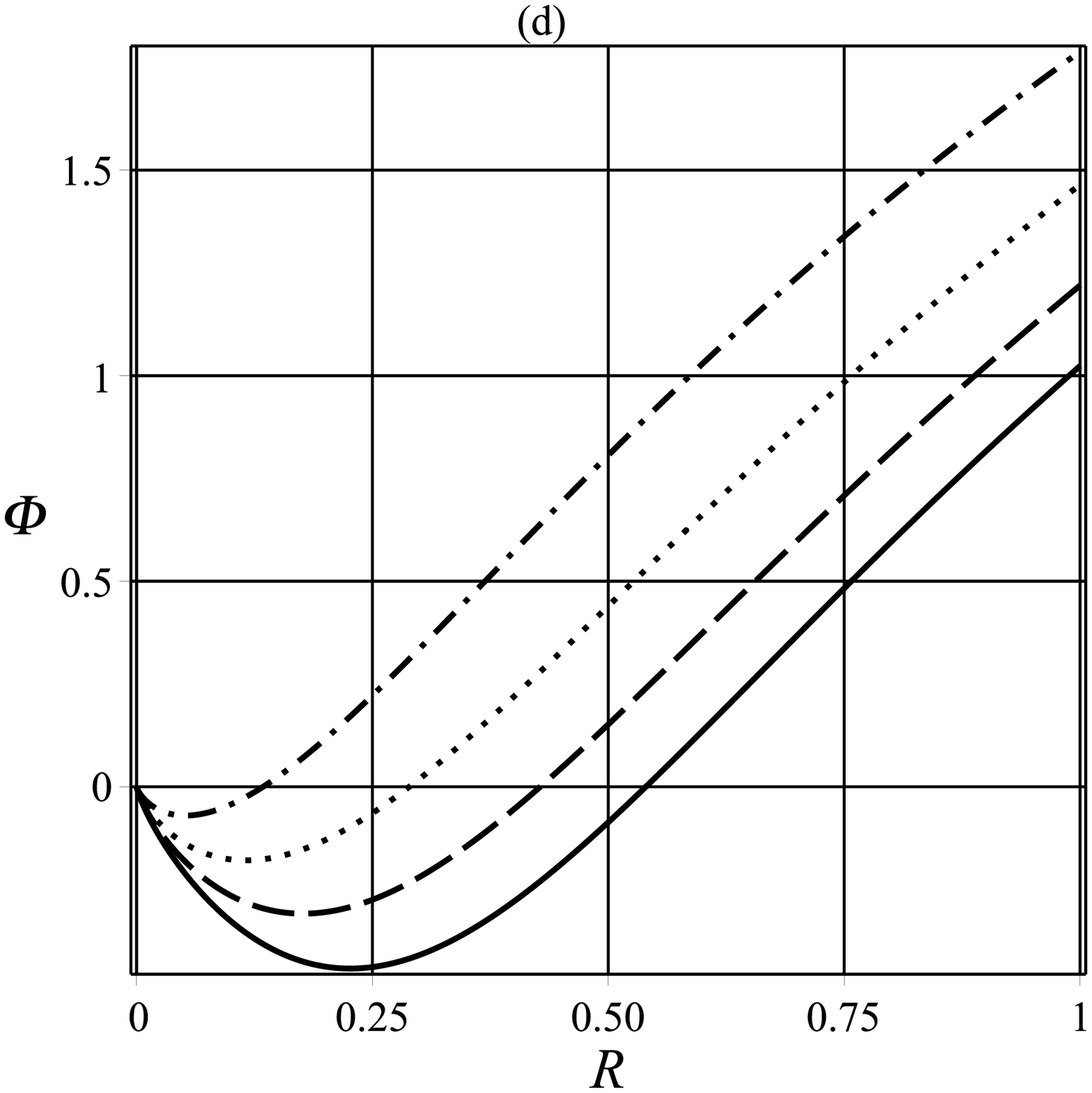}\includegraphics[width=50 mm]{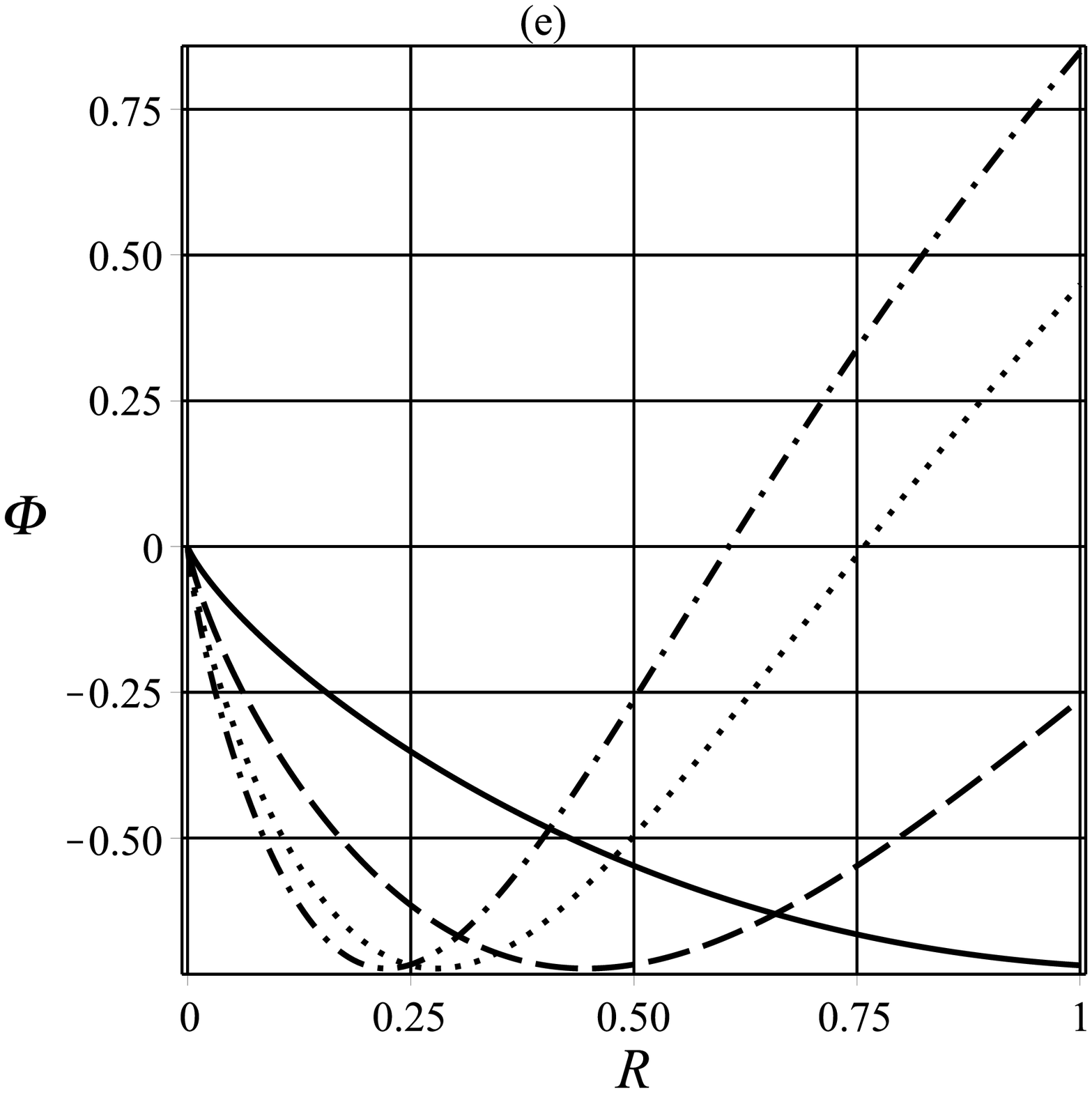}\includegraphics[width=50 mm]{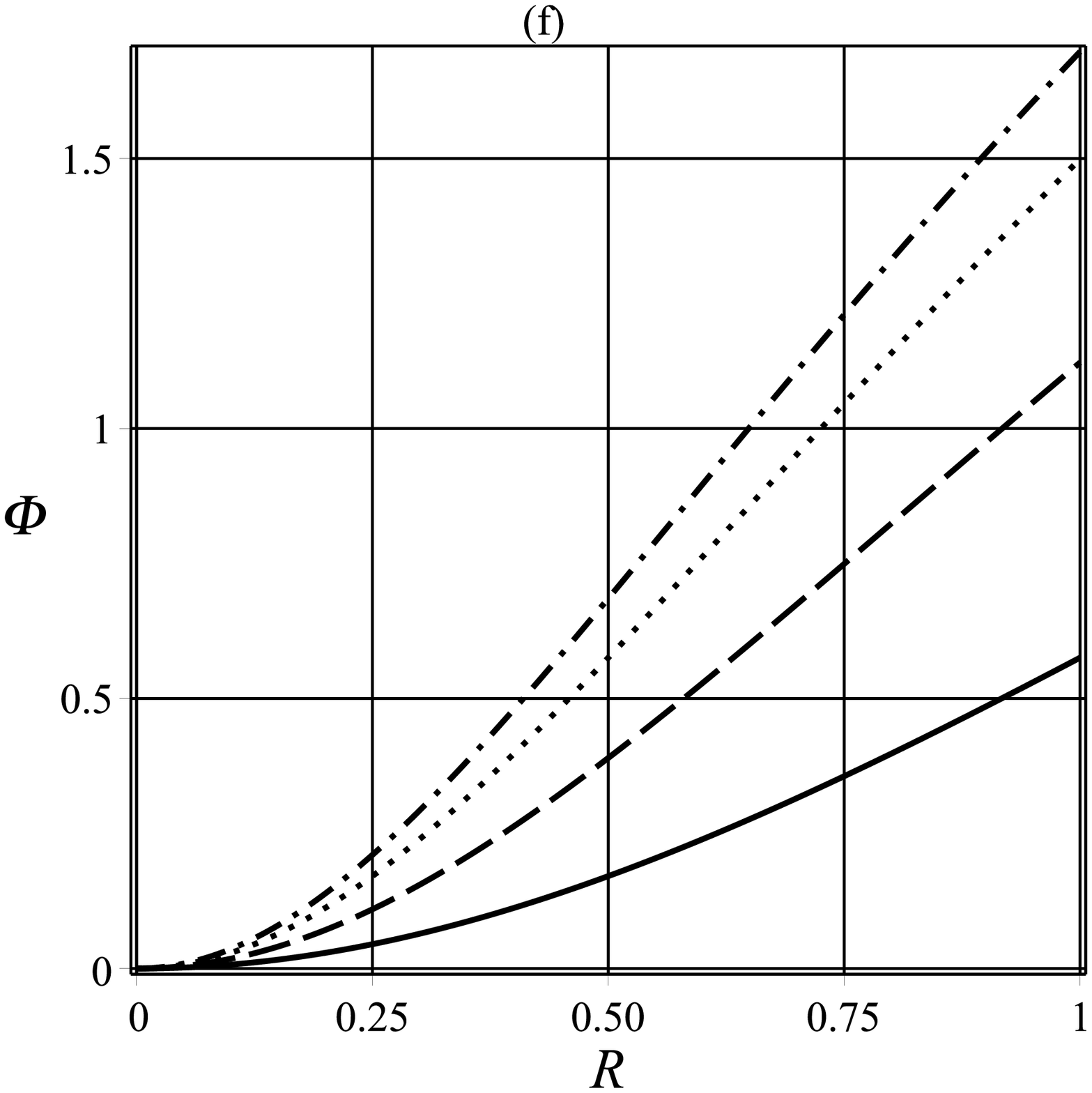}
 \end{array}$
 \end{center}
\caption{The scalar field $\Phi$ versus $R$. (a) $\alpha=\beta=\gamma=0.5$; $n=3$ (solid), $n=4$ (dash), $n=5$ (dot), $n=6$ (dash dot). (b) $\alpha=\beta=0.5$, $n=3$; $\gamma=0$ (solid), $\gamma=0.2$ (dash), $\gamma=0.5$ (dot), $\gamma=1$ (dash dot). (c) $\alpha=\gamma=0.5$, $n=3$; $\beta=0$ (solid), $\beta=0.2$ (dash), $\beta=0.5$ (dot), $\beta=1$ (dash dot). (d) $\beta=\gamma=0.5$, $n=3$; $\alpha=0$ (solid), $\alpha=0.2$ (dash), $\alpha=0.5$ (dot), $\alpha=1$ (dash dot). (e) $\alpha=\beta=0$, $n=3$; $\gamma=0.2$ (solid), $\gamma=0.5$ (dash), $\gamma=0.8$ (dot), $\gamma=1$ (dash dot). (f) $\alpha=\gamma=0$, $n=3$; $\beta=0.2$ (solid), $\beta=0.5$ (dash), $\beta=0.8$ (dot), $\beta=1$ (dash dot).}
 \label{fig1}
\end{figure}

It is possible to do the following conformal transformation of the metric \cite{9312008},
\begin{equation}\label{S1}
\tilde{g}_{\mu \nu} = f^{\prime}(R)g_{\mu \nu} = (1 + (2\alpha+\gamma) R + 2 \gamma R \ln \beta R+n\beta R^{n-1}) g_{\mu \nu}.
\end{equation}
In that case the action given by the equation (\ref{M1}) with $\pounds_m = 0$ written as,
\begin{equation}\label{S2}
S =  \int d^4 x \sqrt{-g} \left[\frac{1}{2\kappa^2} \tilde{R} -
\frac{1}{2} \tilde{g}^{\mu \nu} \nabla_{\mu} \Phi \nabla_{\nu}\Phi -
V(\phi) \right ],
\end{equation}
where $\nabla_{\mu}$ is the covariant derivative, and $\tilde{R}$ is determined using the conformal metric in the equation (\ref{S1}). We can obtain the scalar field $\Phi$
as follow,
\begin{equation}\label{S3}
\Phi = \sqrt{\frac{3}{2}} \ln(1 + (2\alpha+\gamma) R + 2 \gamma R \ln{\gamma R}+n\beta R^{n-1}).
\end{equation}
In the plots of the Fig. \ref{fig1} we can see behavior of the scalar field for various values of parameters. In the Fig. \ref{fig1} (a) we can see behavior of the scalar field for different $n$. It is illustrated that, evolution of $\Phi$ for larger $n$ is faster than the cases with lower $n$. In the Fig. \ref{fig1} (b) we can see that the scalar field $\Phi$ is linear for $\gamma=0$. Fig. \ref{fig1} (c) tells that increasing $\beta$ parameter, as well as $\alpha$ parameter (see Fig. \ref{fig1} (d)) increases value of the scalar field.\\
Two last figures (e) and (f) show effect of $R^{n}$ and logarithmic corrections separately. Later, we will fix parameters using slow-roll condition.

\begin{figure}[h!]
 \begin{center}$
 \begin{array}{cccc}
\includegraphics[width=60 mm]{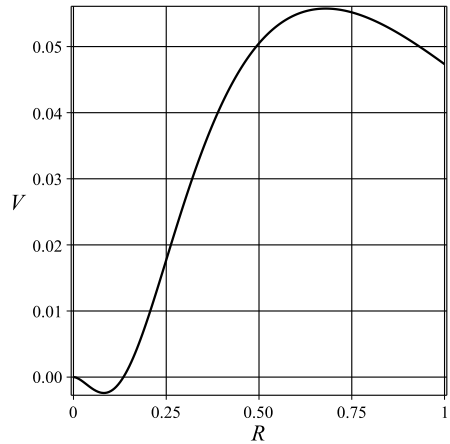}
 \end{array}$
 \end{center}
\caption{The The scalar potential $V$ versus $R$ for $\alpha=1$, $\beta=0.5$, $\gamma=1$, and $n=5$.}
 \label{fig2}
\end{figure}

Then, we can obtain the potential $V$ as follow,
\begin{equation}\label{S4}
V = \frac{\gamma R^{2}\ln(\gamma R)+\beta(n-1)R^{n}+R^{2}(\alpha+\gamma)}{2 (1 + (2\alpha+\gamma) R + 2 \gamma R \ln{\gamma R}+n\beta R^{n-1})^{2}}.
\end{equation}
We can see that for appropriate choice of parameters there is at least a minimum and a maximum for the potential. Typical behavior of the potential plotted in the Fig. \ref{fig2}

\section{Slow-roll parameters}
\begin{figure}[h!]
 \begin{center}$
 \begin{array}{cccc}
\includegraphics[width=50 mm]{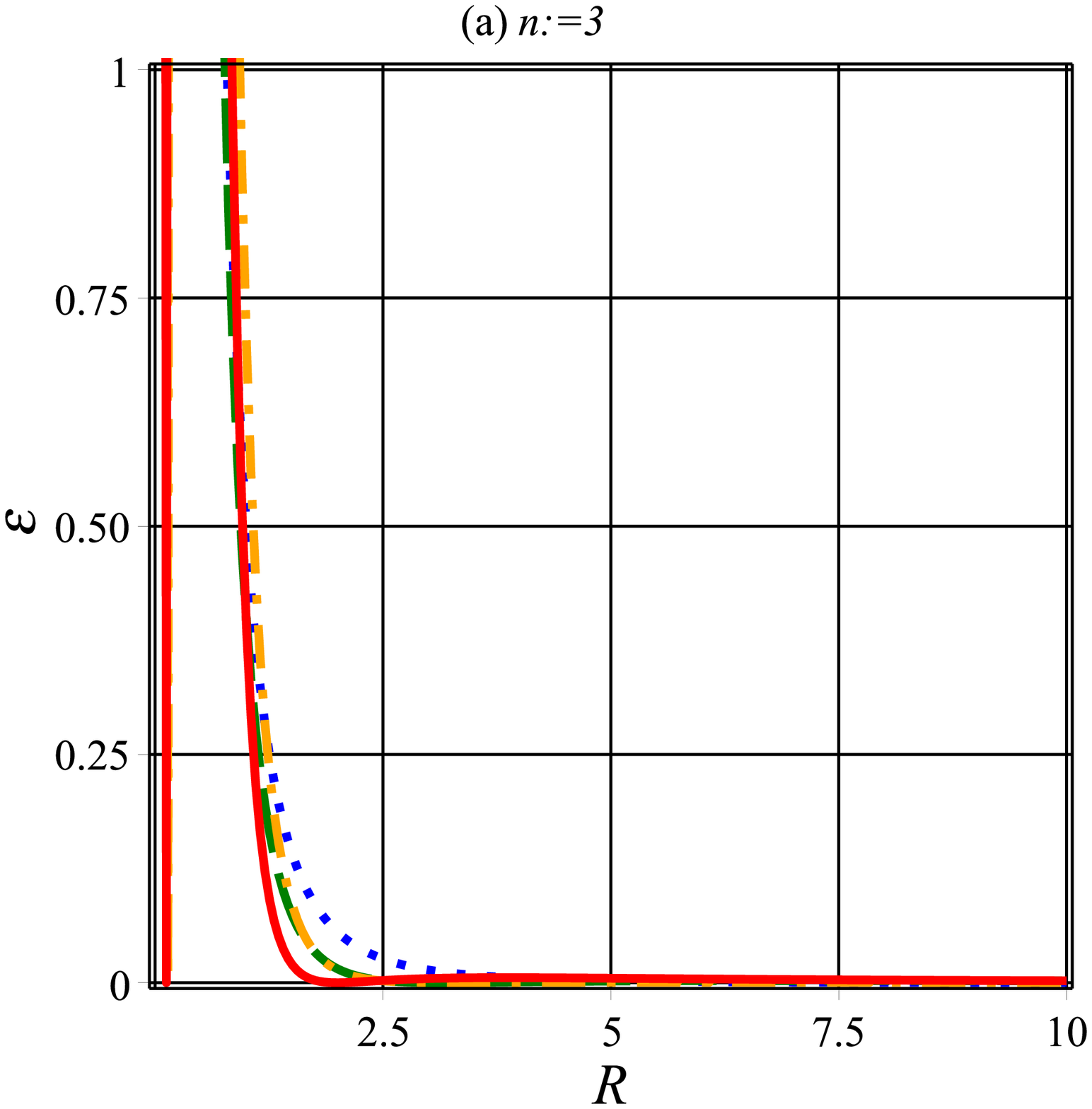}\includegraphics[width=50 mm]{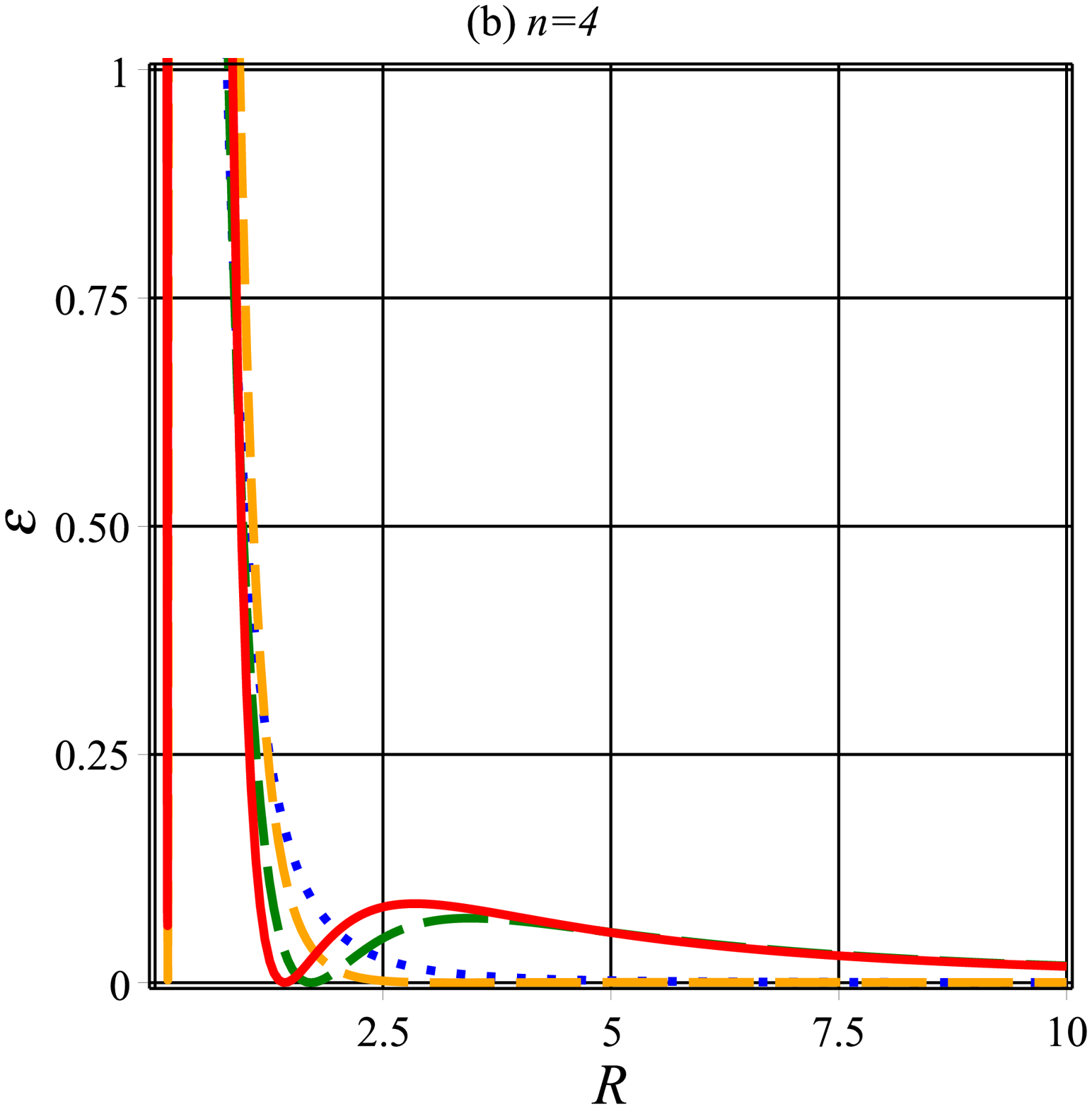}\includegraphics[width=50 mm]{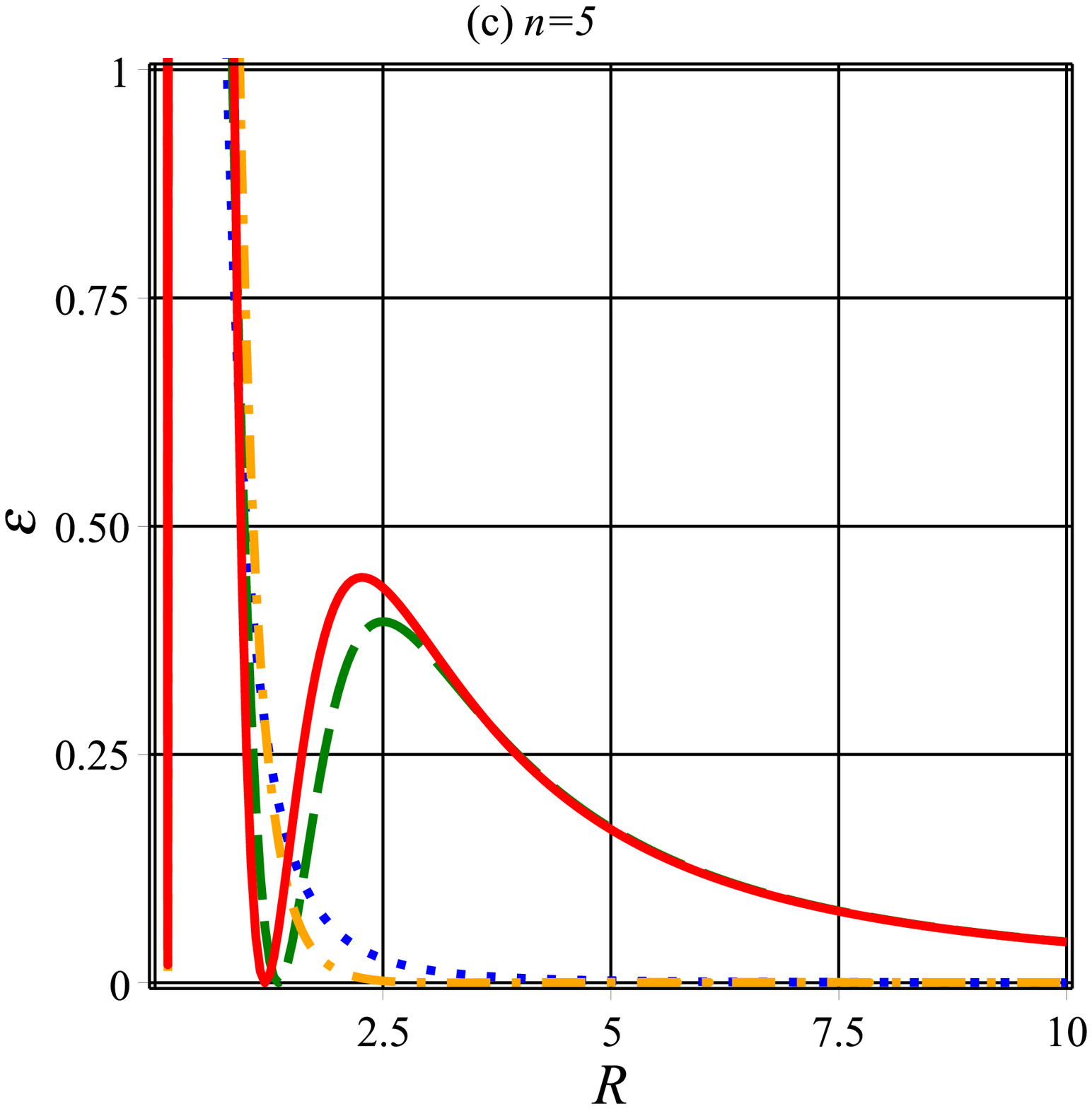}
 \end{array}$
 \end{center}
\caption{$\varepsilon$ versus $R$ for $\alpha=0.5$. (a) $n=3$, (b) $n=4$, (c) $n=5$. $\beta=0$, $\gamma=0$ (blue dot). $\beta=0.1$, $\gamma=0$ (green dash). $\beta=0$, $\gamma=0.3$ (orange dash dot). $\beta=0.1$, $\gamma=0.3$ (red solid).}
 \label{fig3}
\end{figure}

Having scalar potential help us to obtain some important cosmological parameters which constructed using the slow-roll parameters. The slow-roll parameters are given by,
\begin{equation}\label{CP3}
\varepsilon = \frac{1}{2} \left (\frac{V^{\prime}}{V}\right)^{2},
\end{equation}
and
\begin{equation}\label{CP3-2}
\eta =\frac{1}{2} \frac{V^{\prime \prime}}{V},
\end{equation}
where $8\pi G=1$ is used for simplicity.\\
For the slow-roll approximation, the crucial conditions are $\varepsilon \ll 1$ and $ \eta \ll 1 $. One can obtains the
slow-roll parameters expressed through the curvature from the equations (\ref{S4}), (\ref{CP3}) and (\ref{CP3-2}). We can write the following analytical expression of $\varepsilon$,
\begin{equation}\label{CP4}
\varepsilon =\frac{[(\alpha+\frac{3}{2}\gamma)R^{2}+\frac{1}{2}n(n-1)\beta R^{n}+R^{2}\gamma\ln{\gamma R}]^{2}
(\gamma R^{2}-R+(n-2)\beta R^{n})^{2}}{2R^{2}[(\frac{1}{2}+(\alpha+\frac{\gamma}{2})R)R+\frac{1}{2}\beta nR^{2}+\gamma R^{2}\ln{\gamma R}]^{2}
((\alpha+\gamma)R^{2}+(n-1)\beta R^{n}+\gamma R^{2}\ln{\gamma R})^{2}}.
\end{equation}
In the Fig. \ref{fig3} we can check slow-roll condition $\varepsilon\ll1$ for different values of parameters. We can see that in the case of without correction ($\beta=\gamma=0$) the condition $\varepsilon\ll1$ satisfied only for $R\gg1$ with any value of $n$. Effect of $R^{n}$ corrections ($\beta\neq0$, $\gamma=0$) illustrated by green dashed lines. For the cases of $n\geq4$ we have two regions where $\varepsilon\ll1$ as $R\gg1$ and $R\approx1$, while in the case of $n=3$ we have slow-roll condition only for $R\gg1$. On the other hand the effect of logarithmic correction  ($\beta=0$, $\gamma\neq0$) illustrated by orange dash dotted lines. In that case we have two regions where $\varepsilon\ll1$ as $R\gg1$ and $R\approx1$ for any $n$. Finally both corrections ($\beta\neq0$, $\gamma\neq0$) yield to the red solid lines. In that case slow-roll condition $\varepsilon\ll1$ satisfied in three different regions as $R\approx1$, $R\ll1$ and $R\gg1$ for $n\geq4$, while $R\ll1$ and $R\gg1$ for $n=3$.\\
Above discussion suggested that presence of both corrections (logarithmic and $R^{2}$) is necessary to satisfy slow-roll condition $\varepsilon\ll1$ for any arbitrary $n$ with finite $R$ ($R\leq1$), because the end of inflation $\varepsilon\approx\eta\approx1$ happen at $R\approx1$.

\begin{figure}[h!]
 \begin{center}$
 \begin{array}{cccc}
\includegraphics[width=60 mm]{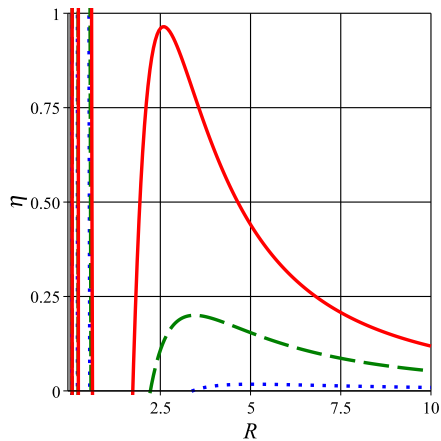}
 \end{array}$
 \end{center}
\caption{$\eta$ versus $R$ for $\alpha=0.5$, $\beta=0.1$, $\gamma=0.3$ and $n=3$ (blue dot), $n=4$ (green dash), $n=5$ (red solid).}
 \label{fig4}
\end{figure}

We can find that above values of parameters satisfy other slow-roll condition $ \eta \ll 1 $. In the Fig. \ref{fig4} we draw typical behavior of $\eta$ and see that we have $\eta\ll1$ at $R\leq1$.

\begin{figure}[h!]
 \begin{center}$
 \begin{array}{cccc}
\includegraphics[width=60 mm]{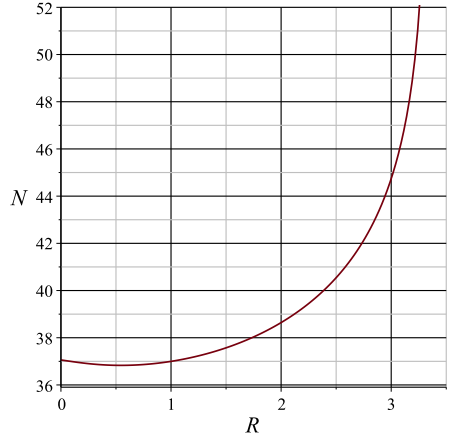}
 \end{array}$
 \end{center}
\caption{$N$ versus $R$ for $\alpha=0.5$, $\beta=0.00005$, $\gamma=0.3$ and $n=3$.}
 \label{fig5}
\end{figure}

The age of the inflation can be obtained by calculating the e-fold number,
\begin{equation}\label{CP6}
N_{e}=\int^{R_{0}} _{R_{end}}\frac{V(\Phi^{\prime})^{2}}{V^{\prime}}dR,
\end{equation}

where $R_{0}$ denotes present day Ricci scalar and $R _{end}$ corresponds to the time of the end of
inflation when $\varepsilon $ or $|\eta| $ are close to $1$ so from the Fig \ref{fig3} we can say $R _{end}\approx1$ as discussed above. Again we find that presence of both corrections are crucial to have $50\leq N\leq60$. For example in the case of $n=3$ with $\alpha$ of order unity, $\gamma=0.3$ and infinitesimal $\beta$ we can obtain $N=52$ at $R\approx3.5$ as present day value of Ricci scalar (see Fig. \ref{fig5}).

\begin{figure}[h!]
 \begin{center}$
 \begin{array}{cccc}
\includegraphics[width=60 mm]{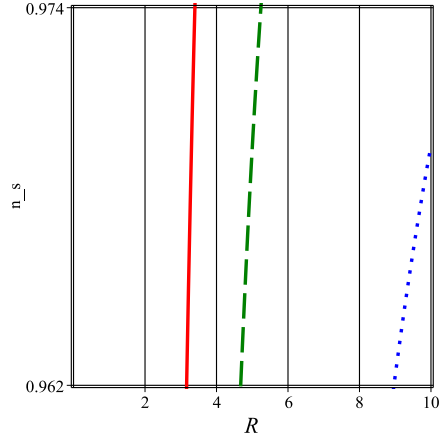}
 \end{array}$
 \end{center}
\caption{$n_{s}$ versus $R$ for $\alpha=1$, $\beta=0.1$, $\gamma=0.3$ and $n=3$ (red solid), $n=4$ (green dash), $n=5$ (blue dot).}
 \label{fig6}
\end{figure}

Having $\varepsilon$ and $\eta$ give us two important cosmological parameters. The index of the
scalar spectrum is given by,
\begin{equation}\label{CP7}
n_{s} = 1 - 6 \varepsilon + 2 \eta.
\end{equation}
Our results illustrated by the Fig. \ref{fig6} which suggest $n=3$ is more agreement with observations while other parameters should be fixed as before. Therefore, both logarithmic and polynomial corrections are necessary to have agreement with observations.

\begin{figure}[h!]
 \begin{center}$
 \begin{array}{cccc}
\includegraphics[width=60 mm]{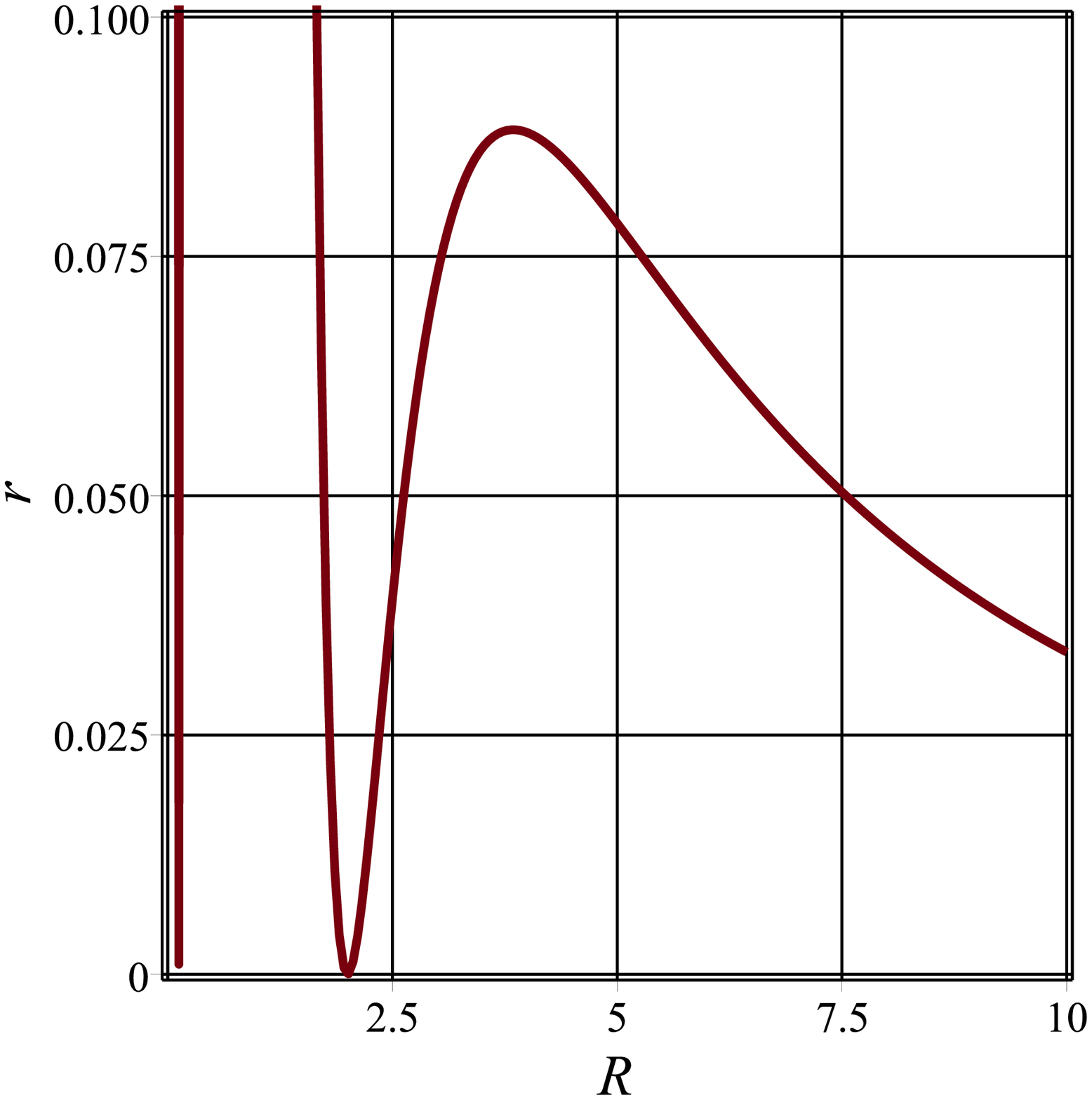}
 \end{array}$
 \end{center}
\caption{$r$ versus $R$ for $\alpha=1$, $\beta=0.1$, $\gamma=0.3$ and $n=3$.}
 \label{fig7}
\end{figure}

Also, the tensor-to-scalar ratio is defined by,
\begin{equation}\label{CP8}
r=16 \varepsilon.
\end{equation}
In the Fig. \ref{fig7} we show that observational bound $r < 0.09$, denoted in the introduction, satisfied.

\section{Critical points and stability}
There are several ways to study stability of a cosmological model. Here, we follow method of the Ref. \cite{0612180} to find critical points and investigate stability of the model.
In order to investigate critical points of equations of motion, it is useful to introduce the following dimensionless parameters,
\begin{equation}\label{CPS1}
x_{1}=-\frac{2\alpha+\beta n(n-1)R^{n-2}+\gamma(1+4\ln{\gamma R})}{1+\beta n R^{n-1}+(2\alpha+\gamma)R+2\gamma R\ln{\gamma R}}\frac{\dot{R}}{H},
\end{equation}
where $H$ is Hubble expansion parameter and dot denote time derivative,
\begin{equation}\label{CPS2}
x_{2}=-\frac{1}{6H^{2}}\frac{R+\alpha R^{2}+\beta R^{n}+\gamma R^{2}ln{\gamma R}}{1+\beta n R^{n-1}+(2\alpha+\gamma)R+2\gamma R\ln{\gamma R}},
\end{equation}
and $x_{3}=\frac{R}{6H^{2}}$. For example the deceleration and effective equation of state parameters can express in terms of $x_{3}$ as $q=1-x_{3}$ and $\omega_{eff}=-\frac{1}{3}(2x_{3}-1)$. Using above parameters we can specify some points $P=(x_{1}, x_{2}, x_{3})$. Moreover, the following identity obtained in terms of the above parameters,
\begin{equation}\label{CPS3}
\Omega=1-x_{1}-x_{2}-x_{3}.
\end{equation}
Also, there are two interesting parameters as follows,
\begin{equation}\label{CPS4}
m=\frac{ 2 \alpha R +\beta n(n-1)R^{n-1}+2\gamma R\ln{\gamma R}+3\gamma R}{1+ (2\alpha+\gamma) R + 2\gamma R \ln(\gamma R)+\beta n R^{n-1}},
\end{equation}
and
\begin{equation}\label{CPS5}
s =- \frac{1+ (2\alpha+\gamma) R + 2\gamma R \ln(\gamma R)+\beta n R^{n-1}}{ 1 + \alpha R +  \beta R^{n-1}+\gamma R \ln(\gamma R)}.
\end{equation}
It is interesting to note that $s=\frac{x_{3}}{x_{2}}$. It means that
\begin{equation}\label{CPS6}
R=- \frac{f}{f^{\prime}}\frac{x_{3}}{x_{2}},
\end{equation}
therefore we can express $m$ in terms of $s$,
\begin{equation}\label{CPS7}
m=- \frac{ff^{\prime\prime}}{f^{\prime2}}\frac{x_{3}}{x_{2}}=- \frac{ff^{\prime\prime}}{f^{\prime2}}s,
\end{equation}
and the critical points of the system can obtain by investigation of the function $m(s)$, which shows the
deviation from the $\Lambda$CDM model. Below we list and discuss about some critical points without radiation.
\subsection{de-Sitter}
The first critical point $P_{1}=(0, -1, 2)$ with $q=\omega_{eff}=-1$ and $\Omega=0$ corresponds to de-Sitter solution. This point mimics a cosmological constant. It is obvious that $s=-2$, then combination of it with the relation (\ref{CPS6}) yields to the condition given by the equation (\ref{CCC2}). Therefore, the point $P_{1}$ corresponds to the constant curvature solutions introduced in section 3. $x_{3}=2$ tells that $R=12H$, and we can find that $\dot{H}=0$.\\
In the special case of $n=3$ we find the solution (\ref{n=3}). In that case, if we assume $4\beta=-\gamma^{2}$, then $m=1$. It is indeed similar to the simplest case of $\beta=\gamma=0$ which is uncorrected model.

\subsection{kinetic epoch}
The second and third critical points $P_{2,3}=(\pm1, 0, 0)$ yields to $q=1$ and $\omega_{eff}=\frac{1}{3}$. For positive and negative sign we have $P_{2}$ and $P_{3}$ with $\Omega_{+}=0$ and $\Omega_{-}=2$ corresponds to a purely kinetic epoch and a field-matter dominated epoch respectively introduced by Amendola et al. \cite{0011243,9908023,0303228}. As pointed by the Ref. \cite{0612180} for several models like logarithmic, power-law and polynomial function, one need to apply de l'Hopital rule to vanish imaginary picture of $\frac{x_{3}}{x_{2}}=\frac{0}{0}$, and all yield to results of the Ref. \cite{0612180}.

\subsection{scaling solutions}
\begin{figure}[h!]
 \begin{center}$
 \begin{array}{cccc}
\includegraphics[width=60 mm]{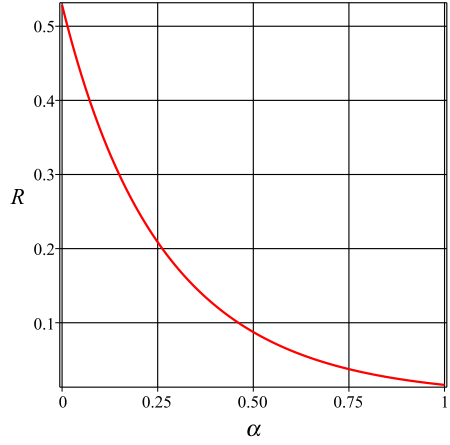}
 \end{array}$
 \end{center}
\caption{$R$ versus $\alpha$ for $n=3$, $\beta=0$, $\gamma=0.3$ (solid red), $\gamma=0.5$ (dotted blue), $\gamma=0.7$ (dashed green).}
 \label{fig8}
\end{figure}

\begin{figure}[h!]
 \begin{center}$
 \begin{array}{cccc}
\includegraphics[width=45 mm]{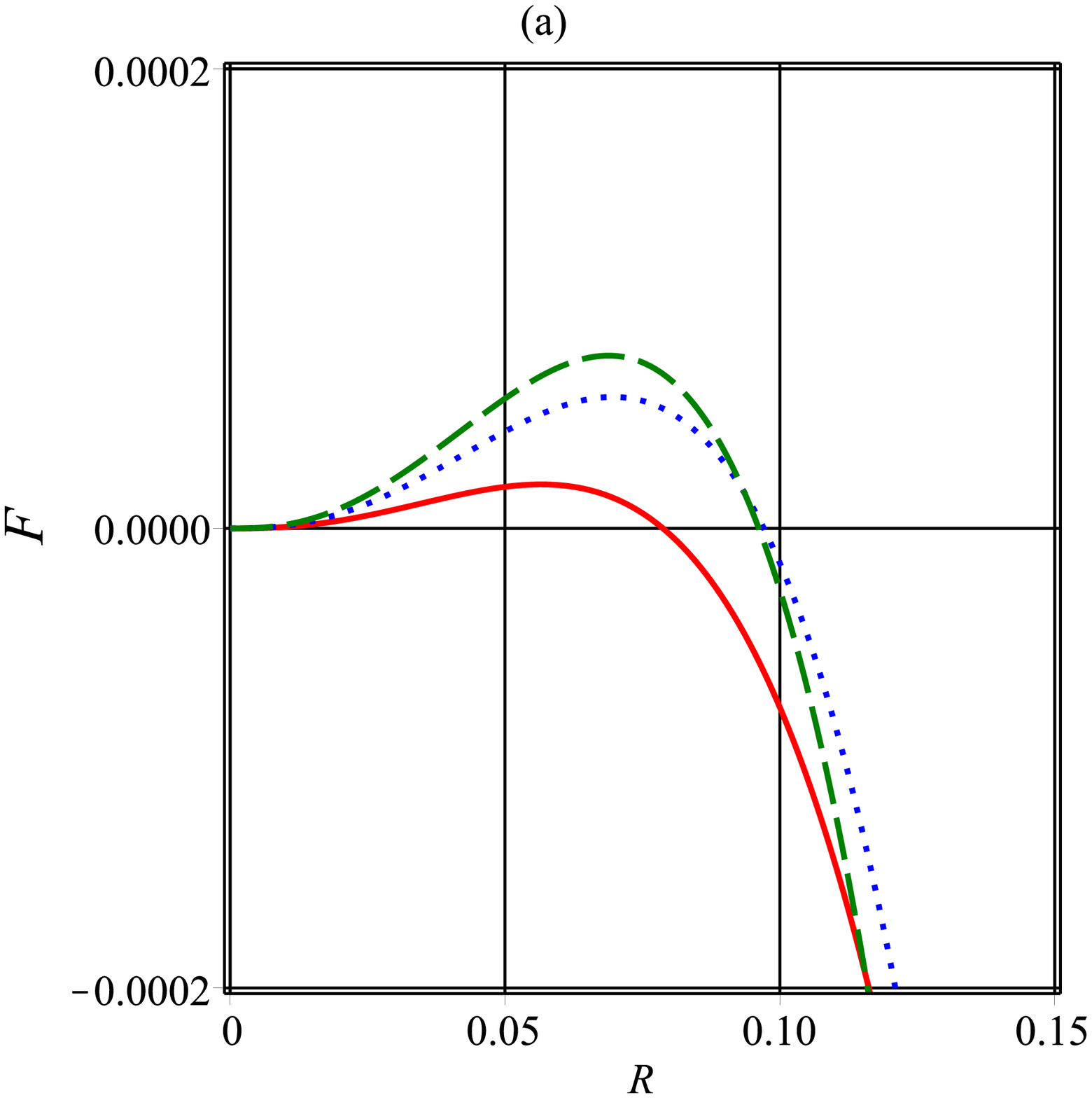}\includegraphics[width=45 mm]{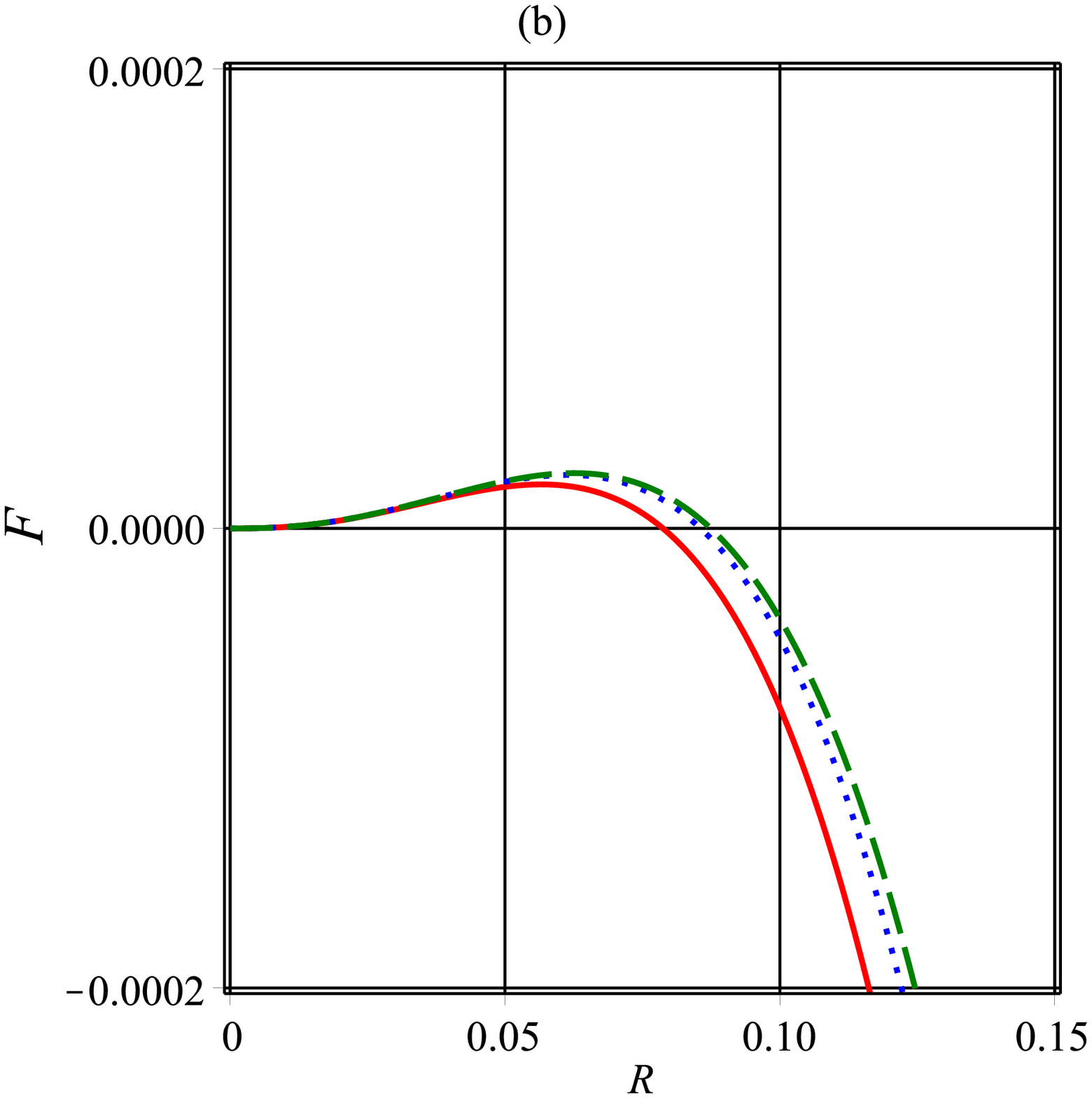}\includegraphics[width=45 mm]{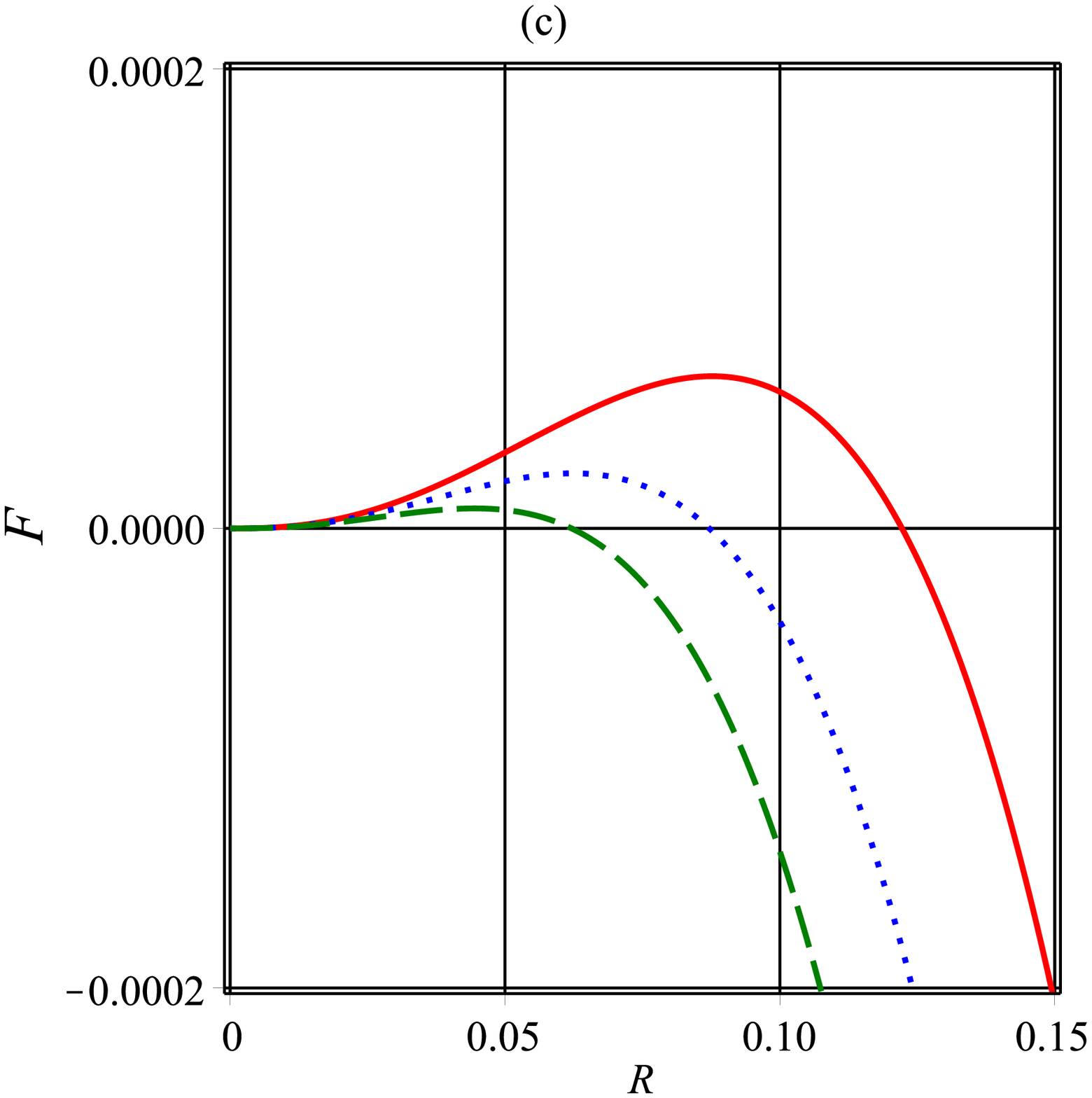}
 \end{array}$
 \end{center}
\caption{$F$ versus $R$ for $\beta=0.1$. (a) $\alpha=0.5$, $n=3$, $\gamma=0.3$ (solid red), $\gamma=0.5$ (dotted blue), $\gamma=0.7$ (dashed green). (b) $\alpha=0.5$, $\gamma=0.3$, $n=3$ (solid red), $n=4$ (dotted blue), $n=5$ (dashed green). (c) $\gamma=0.3$, $n=3$, $\alpha=0.4$ (solid red), $\alpha=0.5$ (dotted blue), $\alpha=0.6$ (dashed green).}
 \label{fig9}
\end{figure}

For the other critical point,
\begin{equation}\label{CPS8}
P_{4}= \left(\frac{3m}{1+m},-\frac{1+4m}{2(1+m)^{2}},\frac{1+4m}{2(1+m)}\right),
\end{equation}
one can find,
\begin{eqnarray}\label{CPS9}
q&=&\frac{1-2m}{2(1+m)},\nonumber\\
\omega_{eff}&=&-\frac{m}{1+m},\nonumber\\
\Omega&=&1-\frac{m(7+10m)}{2(1+m)^{2}}.
\end{eqnarray}
Also it is clear that $s=-(m+1)$.\\
In the special case of $m=0$ where $x _{3}=\frac{1}{2}$, $ r =  -1$, $q=\frac{1}{2}$, $\omega_{eff} = 0$, and $\Omega=1$ we recover standard mater era with $a = a_{0} t^{\frac{2}{3}}$. Then, we have a viable matter dominated epoch prior to late-time acceleration.\\
Independent of correction parameters, general behavior of above quantities has been studied by the Ref. \cite{0612180}. Here we would like to investigate effect of $\beta$ and $\gamma$ (polynomial and logarithmic effect respectively). Combination of the equations (\ref{CPS4}), (\ref{CPS5}) and (\ref{CPS8}) suggest the following equation,
\begin{eqnarray}\label{CPS9-1}
&-&\left[\gamma+\beta(n-2)^{2}R^{n-1}\right]\gamma\ln{\gamma R}\nonumber\\
&+&\left[2(2\alpha+\gamma)n-\alpha n^{2}-4(\alpha+\gamma)\right]\beta R^{n-1}\nonumber\\
&-&\beta(n-1)^{2}R^{n-2}+\gamma^{2} R-(\alpha+2\gamma)=F=0.
\end{eqnarray}
It is obvious that $\beta=\gamma=0$ is forbidden unless $\alpha=0$, so we find that in presence of $\alpha$, at least one of $\beta$ or $\gamma$ should exist. We will show that logarithmic correction is necessary to have positive $R$. Therefore, we consider two special case of $\beta=0$, $\gamma\neq0$ and $\beta\neq0$, $\gamma=0$ both for $n=3$.\\
In the first case we set $n=3$ and $\gamma=0$, so the equation (\ref{CPS9-1}) has the following roots,
\begin{equation}\label{CPS9-2}
R=\frac{-2\beta\pm\sqrt{4\beta^{2}-\beta\alpha^{2}}}{\alpha\beta}<0,
\end{equation}
which are clearly negative for any positive $\alpha$ and $\beta$. We have similar problem for higher $n$. Hence we seek other solutions with $\beta=0$ in the equation (\ref{CPS9-1}) which gives,
\begin{equation}\label{CPS9-3}
R=\frac{1}{\gamma}e^{\frac{\alpha+2\gamma}{\gamma}}e^{-LW(-e^{-frac{\alpha+2\gamma}{\gamma}})},
\end{equation}
where $LW$ stands for Lambert function. It is clear that with $\beta=0$ value of $n$ is not important. In the Fig. \ref{fig8} we draw $R$ to see that $\gamma$ and $\alpha$ reduce value of $R$, but with selected values of them we have positive $R$. In this case we can find $m$ as infinitesimal negative parameter.\\
In the Fig. \ref{fig9} we consider general form of the equation (\ref{CPS9-1}) and draw $F$ in terms of $R$ to find roots of the equation. In the general case, we can see that appropriate choice of parameters, as before, yield to at least two roots.

\subsection{curvature-dominated point}
The last critical point given by,
\begin{equation}\label{CPS10}
P_{5}= \left(\frac{2(1-m)}{1+2m},-\frac{1-4m}{m(1+2m)},-\frac{(1-4m)(1+m)}{m(1+2m)}\right),
\end{equation}
one can find,
\begin{eqnarray}\label{CPS11}
q&=&\frac{1-2m(1+m)}{m(1+2m)},\nonumber\\
\omega_{eff}&=&\frac{2-5m-6m^{2}}{3m(1+2m)},\nonumber\\
\Omega&=&0.
\end{eqnarray}
Also, it is clear that $s=-(m+1)$ exactly similar to the point $P_{4}$. Therefore, we have similar description with Fig. \ref{fig8} and Fig. \ref{fig9}. So, in order to have stability, presence of at least logarithmic correction is necessary.\\

\section{Conclusion}
In this paper, we studied an inflationary model based on a specific $f(R)$ gravity theory with the choice of $f(R) = R+\alpha R^{2}+\beta R^{n}+\gamma R^{2}\ln{\gamma R}$. We concluded that presence of, at least, logarithmic correction is necessary to have a successful model. However, polynomial correction ($\beta\neq0$) with appropriate $\beta$ some time help to give a result in agreement with observational data.
In particular, the model parameters are examined by using the recent Planck 2015 results for the tensor to scalar ratio. The evolution of the scalar field and scalar potential also investigated. We discussed about the equation of state parameter for the universe through the autonomous system analysis. So, we considered some critical points of the model and discussed about them. The mathematical investigations might be useful for the future studies. For example it is interesting to consider another $f(R)$ gravity models.

\end{document}